\documentclass[aps, prl, superscriptaddress, showpacs, preprintnumbers]{revtex4}

\usepackage{graphics}
\usepackage[dvipdfmx]{graphicx}

\begin{document}

\title{\quad\\[1.0cm] 
First model-independent Dalitz analysis of $B^0 \to DK^{*0}$, $D\to K_S^0\pi^+\pi^-$ decay}


\noaffiliation
\affiliation{University of the Basque Country UPV/EHU, 48080 Bilbao}
\affiliation{Beihang University, Beijing 100191}
\affiliation{University of Bonn, 53115 Bonn}
\affiliation{Budker Institute of Nuclear Physics SB RAS and Novosibirsk State University, Novosibirsk 630090}
\affiliation{Faculty of Mathematics and Physics, Charles University, 121 16 Prague}
\affiliation{Chonnam National University, Kwangju 660-701}
\affiliation{University of Cincinnati, Cincinnati, Ohio 45221}
\affiliation{Deutsches Elektronen--Synchrotron, 22607 Hamburg}
\affiliation{University of Florida, Gainesville, Florida 32611}
\affiliation{Justus-Liebig-Universit\"at Gie\ss{}en, 35392 Gie\ss{}en}
\affiliation{Gifu University, Gifu 501-1193}
\affiliation{SOKENDAI (The Graduate University for Advanced Studies), Hayama 240-0193}
\affiliation{Gyeongsang National University, Chinju 660-701}
\affiliation{Hanyang University, Seoul 133-791}
\affiliation{University of Hawaii, Honolulu, Hawaii 96822}
\affiliation{High Energy Accelerator Research Organization (KEK), Tsukuba 305-0801}
\affiliation{IKERBASQUE, Basque Foundation for Science, 48013 Bilbao}
\affiliation{Indian Institute of Technology Bhubaneswar, Satya Nagar 751007}
\affiliation{Indian Institute of Technology Madras, Chennai 600036}
\affiliation{Indiana University, Bloomington, Indiana 47408}
\affiliation{Institute of High Energy Physics, Chinese Academy of Sciences, Beijing 100049}
\affiliation{Institute of High Energy Physics, Vienna 1050}
\affiliation{INFN - Sezione di Torino, 10125 Torino}
\affiliation{Institute for Theoretical and Experimental Physics, Moscow 117218}
\affiliation{J. Stefan Institute, 1000 Ljubljana}
\affiliation{Kanagawa University, Yokohama 221-8686}
\affiliation{Institut f\"ur Experimentelle Kernphysik, Karlsruher Institut f\"ur Technologie, 76131 Karlsruhe}
\affiliation{Kennesaw State University, Kennesaw GA 30144}
\affiliation{King Abdulaziz City for Science and Technology, Riyadh 11442}
\affiliation{Department of Physics, Faculty of Science, King Abdulaziz University, Jeddah 21589}
\affiliation{Korea Institute of Science and Technology Information, Daejeon 305-806}
\affiliation{Korea University, Seoul 136-713}
\affiliation{Kyungpook National University, Daegu 702-701}
\affiliation{\'Ecole Polytechnique F\'ed\'erale de Lausanne (EPFL), Lausanne 1015}
\affiliation{Faculty of Mathematics and Physics, University of Ljubljana, 1000 Ljubljana}
\affiliation{Luther College, Decorah, Iowa 52101}
\affiliation{University of Maribor, 2000 Maribor}
\affiliation{Max-Planck-Institut f\"ur Physik, 80805 M\"unchen}
\affiliation{School of Physics, University of Melbourne, Victoria 3010}
\affiliation{Moscow Physical Engineering Institute, Moscow 115409}
\affiliation{Moscow Institute of Physics and Technology, Moscow Region 141700}
\affiliation{Graduate School of Science, Nagoya University, Nagoya 464-8602}
\affiliation{Kobayashi-Maskawa Institute, Nagoya University, Nagoya 464-8602}
\affiliation{Nara Women's University, Nara 630-8506}
\affiliation{National Central University, Chung-li 32054}
\affiliation{National United University, Miao Li 36003}
\affiliation{Department of Physics, National Taiwan University, Taipei 10617}
\affiliation{H. Niewodniczanski Institute of Nuclear Physics, Krakow 31-342}
\affiliation{Nippon Dental University, Niigata 951-8580}
\affiliation{Niigata University, Niigata 950-2181}
\affiliation{University of Nova Gorica, 5000 Nova Gorica}
\affiliation{Osaka City University, Osaka 558-8585}
\affiliation{Pacific Northwest National Laboratory, Richland, Washington 99352}
\affiliation{Peking University, Beijing 100871}
\affiliation{University of Pittsburgh, Pittsburgh, Pennsylvania 15260}
\affiliation{University of Science and Technology of China, Hefei 230026}
\affiliation{Soongsil University, Seoul 156-743}
\affiliation{Sungkyunkwan University, Suwon 440-746}
\affiliation{School of Physics, University of Sydney, NSW 2006}
\affiliation{Department of Physics, Faculty of Science, University of Tabuk, Tabuk 71451}
\affiliation{Tata Institute of Fundamental Research, Mumbai 400005}
\affiliation{Excellence Cluster Universe, Technische Universit\"at M\"unchen, 85748 Garching}
\affiliation{Toho University, Funabashi 274-8510}
\affiliation{Tohoku University, Sendai 980-8578}
\affiliation{Earthquake Research Institute, University of Tokyo, Tokyo 113-0032}
\affiliation{Department of Physics, University of Tokyo, Tokyo 113-0033}
\affiliation{Tokyo Institute of Technology, Tokyo 152-8550}
\affiliation{Tokyo Metropolitan University, Tokyo 192-0397}
\affiliation{University of Torino, 10124 Torino}
\affiliation{CNP, Virginia Polytechnic Institute and State University, Blacksburg, Virginia 24061}
\affiliation{Wayne State University, Detroit, Michigan 48202}
\affiliation{Yamagata University, Yamagata 990-8560}
\affiliation{Yonsei University, Seoul 120-749}

 \author{K.~Negishi}\affiliation{Tohoku University, Sendai 980-8578} 
 \author{A.~Ishikawa}\affiliation{Tohoku University, Sendai 980-8578} 
 \author{H.~Yamamoto}\affiliation{Tohoku University, Sendai 980-8578} 

  \author{A.~Abdesselam}\affiliation{Department of Physics, Faculty of Science, University of Tabuk, Tabuk 71451} 
  \author{I.~Adachi}\affiliation{High Energy Accelerator Research Organization (KEK), Tsukuba 305-0801}\affiliation{SOKENDAI (The Graduate University for Advanced Studies), Hayama 240-0193} 
  \author{H.~Aihara}\affiliation{Department of Physics, University of Tokyo, Tokyo 113-0033} 
  \author{S.~Al~Said}\affiliation{Department of Physics, Faculty of Science, University of Tabuk, Tabuk 71451}\affiliation{Department of Physics, Faculty of Science, King Abdulaziz University, Jeddah 21589} 
  \author{D.~M.~Asner}\affiliation{Pacific Northwest National Laboratory, Richland, Washington 99352} 
  \author{V.~Aulchenko}\affiliation{Budker Institute of Nuclear Physics SB RAS and Novosibirsk State University, Novosibirsk 630090} 
  \author{T.~Aushev}\affiliation{Moscow Institute of Physics and Technology, Moscow Region 141700}\affiliation{Institute for Theoretical and Experimental Physics, Moscow 117218} 
% ...
\collaboration{The Belle Collaboration}

\begin{abstract}
We report a measurement of the amplitude ratio $r_S$ of $B^0 \to D^0K^{*0}$ and $B^0 \to \bar{D^0}K^{*0}$ decays
with a Dalitz analysis of $D\to K_S^0\pi^+\pi^-$ decays,
for the first time using a model-independent method.
We set an upper limit $r_S < 0.87$ at the 68\% confidence level, using the full data sample of $711~{\rm fb}^{-1}$ corresponding to $772\times10^6$ $B\bar{B}$ pairs collected 
at the $\Upsilon(4S)$ resonance with the Belle detector at the KEKB $e^+e^-$ collider.
This result is obtained from observables 
$x_- = +0.4 ^{+1.0 +0.0}_{-0.6 -0.1} \pm0.0$,
$y_- = -0.6 ^{+0.8 +0.1}_{-1.0 -0.0} \pm0.1$,
$x_+ = +0.1 ^{+0.7 +0.0}_{-0.4 -0.1} \pm0.1$ and	
$y_+ = +0.3 ^{+0.5 +0.0}_{-0.8 -0.1} \pm0.1$,
where $x_\pm = r_S \cos(\delta_S \pm \phi_3)$, $y_\pm = r_S \sin(\delta_S \pm \phi_3)$ and
$\phi_3~(\delta_S)$ is the weak (strong) phase difference between $B^0 \to D^0K^{*0}$ and $B^0 \to \bar{D^0}K^{*0}$.
\end{abstract}


\maketitle

\section{INTRODUCTION}
Determination of parameters of the standard model (SM)
plays an important role in the search for new physics.
In the SM,
the Cabibbo-Kobayashi-Maskawa (CKM) matrix~\cite{CKM} gives a successful description of current all measurements of $CP$ violation.
The $CP$-violating parameters $\phi_1$, $\phi_2$ and $\phi_3$ are the three angles of the most equilateral of the CKM unitarity triangles, of which
$\phi_3 \equiv \arg{(-V_{ud}{V_{ub}}^*/V_{cd}{V_{cb}}^*)}$
is the least accurately determined.
In the usual quark-phase convention,
where the complex phase is negligible in the CKM matrix elements other than $V_{ub}$ and $V_{td}$~\cite{Wolfenstein},
the measurement of $\phi_3$ is equivalent to the extraction of the phase of $V_{ub}$.
To date, $\phi_3$ measurements have been performed mainly with $B$ meson decays into $D^{(*)}K^{(*)}$ final states
~\cite{GLW_Belle_result, Dalitz_Belle_result, GLW_CDF_result, GLW_BaBar_result, ADS_BaBar_result, Dalitz_BaBar_result, ADS_CDF_result, ADS_Belle_result, LHCb_result, BaBar_DKst0},
all of which exploit the interference between the $\bar{D}^{(*)0}$ and $D^{(*)0}$ decaying into a common final state.
In particular, Dalitz analyses of $B^\pm\to D^{(*)}K^{(*)\pm}$, $D\to K_S^0\pi^-+\pi^-$ provide the most precise determination of $\phi_3$.
The Dalitz analysis technique for the measurement of $\phi_3$ was proposed in Ref.~\cite{GGSZ}. 
Belle reported the first $\phi_3$ measurement with the model-independent Dalitz analysis technique in Ref.~\cite{Antn_Mod_Ind_Dalitz},
which exploits a set of measured strong phases instead of relying on a $D$ decay model into a three-body final state.

\begin{figure}[ht]
  \begin{center}
    {\includegraphics[width=0.83\textwidth, bb=0 0 2089 348, clip]{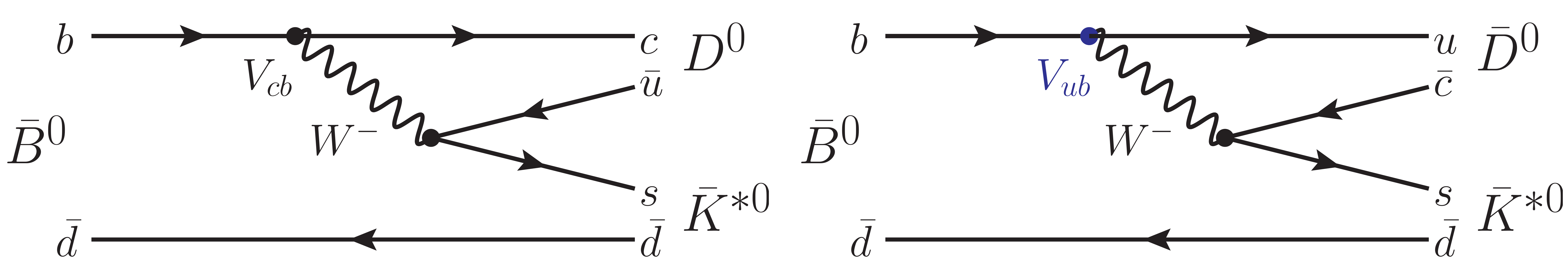}}
    \caption{Diagrams for the $\bar{B}^0 \to D \bar{K}^{*0}$ decay.}
    \label{fig:Dgrm_DKst0_0}
  \end{center}
\end{figure}

In this paper,
we present the first measurement of the amplitude ratio of $B^0 \to D^0K^{*0}$ and $B^0 \to \bar{D^0}K^{*0}$ decays
with a model-independent Dalitz analysis.
We reconstruct $B^0\to DK^{*0}$, with $K^{*0}\to K^+\pi^-$
(Throughout the paper, charge-conjugate processes are implied;
$K^{*0}$ refers to $K^{*}(892)^{0}$
and $D$ refers to either $D^0$ or $\bar{D}^0$ when the $D^0$ flavor is untagged).
Here, the flavor of the $B$ meson is identified by the kaon charge.
Neutral $D$ mesons are reconstructed in the $K_S^0\pi^+\pi^-$ decay mode.
The reconstructed final states are accessible through $b\to c$ and $b\to u$ processes via the diagrams
shown in Fig.~\ref{fig:Dgrm_DKst0_0}.

In this analysis, we use the variables $r_S$, $k$, and $\delta_S$ to parameterize the strong dynamics of the decay. 
These parameters are defined as \cite{ADS_2}
\begin{eqnarray}
  r_S^2 &\equiv& \frac{\Gamma(B^0\to D^0K^+\pi^-)}{\Gamma(B^0\to \bar{D}^0K^+\pi^-)} = \frac{\int {\mathrm d}p A_{b\to u}^2(p)}{\int {\mathrm d}p A_{b\to c}^2(p)}, \\
  k\mathrm{e}^{i\delta_S} &\equiv& \frac{\int {\mathrm d}p A_{b\to c}(p)A_{b\to u}(p)\mathrm{e}^{i\delta(p)}}{\sqrt{\int {\mathrm d}p A_{b\to c}^2(p) \int {\mathrm d}p A_{b\to u}^2(p)}},
  \label{eq:rS_delS_k}
\end{eqnarray}
where the integration is over the $B^0\to DK^+\pi^-$ Dalitz distribution region corresponding to the $K^{*0}$ resonance.
Here, $A_{b\to c}(A_{b\to u})(p)$ is the magnitude of the amplitude for the $b\to c\ (u)$ transition
and $\delta(p)$ is the relative strong phase, where the variable $p$ indicates the position within the $DK^+\pi^-$ Dalitz distribution.
If the $B^0$ decay can be considered as a $DK^{*0}$ two-body decay,
$r_S$ becomes the ratio of the amplitudes for $b\to u$ and $b\to c$ and $k$ becomes $1$.
According to a simulation study using a Dalitz model based on the measurements in Ref.~\cite{BaBar_k},
the value of $k$ is $0.95\pm0.03$ within the phase space of the $DK^{*0}$ resonance.
The value of $r_S$ is expected to be around $0.4$,
which corresponds na\"{i}vely to $\mid V_{ub}V^*_{cs}\mid / \mid V_{cb}V^*_{us}\mid$
but also depends on strong interaction effects.
For $r_S$,
the best experimental value is reported
by LHCb~\cite{r_S_LHCb} as $r_S = 0.240^{+0.055}_{-0.048}$ (different from zero by 2.7 $\sigma$) 
from $B^0\to DK^{*0}$, $D\to K^+K^-$, $\pi^+\pi^-$, $K^\pm\pi^\mp$ decay.

\section{THE MODEL-INDEPENDENT DALITZ ANALYSIS TECHNIQUE}
The amplitude of the $B^0\to DK^{*0}$, $D\to K_S^0\pi^+\pi^-$
decay is a superposition of the $B^0\to \bar{D}^0K^{*0}$ and $B^0\to D^0K^{*0}$ amplitudes
\begin{eqnarray}
A_B(m_+^2,m_-^2) = \bar{A} + r_S e^{i(\delta_S + \phi_3)}A,
	\label{eq:Amp_Inf}
\end{eqnarray}
where $m_+^2$ and $m_-^2$ are the squared invariant masses of the $K_S^0\pi^+$ and $K_S^0\pi^-$ combinations, respectively,
$\bar{A} = \bar{A}(m_+^2,m_-^2)$
is the amplitude of the $B^0\to \bar{D}^0K^{*0},~\bar{D}^0\to K_S^0\pi^+\pi^-$ decay and
$A = A(m_+^2,m_-^2)$
is the amplitude of the $B^0\to D^0K^{*0},~D^0\to K_S^0\pi^+\pi^-$ decay.
In the case of $CP$ conservation in the $D$ decay,
we have 
$A(m_+^2,m_-^2) = \bar{A}(m_-^2,m_+^2)$
as a $CP$ transformation changes $\pi^\pm \to \pi^\mp$,
thus  $m^2_\pm \to m^2_\mp$.
The Dalitz distribution density of the $D$ decay from $B^0\to DK^{*0}$ is given by
\begin{eqnarray}
P_B = \mid A_B \mid^2 = \mid \bar{A} + r_S e^{i(\delta_S + \phi_3)}A \mid^2 = \bar{P} + r_S^2P + 2k\sqrt{P\bar{P}}(x_+C + y_+S),
	\label{eq:Dcy_inf}
\end{eqnarray}
where $P=P(m_+^2,$ $m_-^2) = \mid A \mid^2$, $\bar{P}=\bar{P}(m_+^2,$ $m_-^2) = \mid \bar{A} \mid^2$, and
\begin{eqnarray}
x_+ =r_S\cos(\delta_S + \phi_3), \hspace{20pt} y_+ = r_S\sin(\delta_S + \phi_3).
	\label{eq:x}
\end{eqnarray}
The functions
$C(m_+^2,$ $m_-^2)$ and
$S(m_+^2,$ $m_-^2)$
are the cosine and sine of the strong-phase difference
$\delta_D(m_+^2,$ $m_-^2) = \arg \bar{A} - \arg A$ between the $\bar{D}^0\to K_S^0\pi^+\pi^-$ and $D^0\to K_S^0\pi^+\pi^-$ amplitudes.
Here, we have used the definition of $k$ given in Eq.~(\ref{eq:rS_delS_k}).
The equations for the charge-conjugate mode $\bar{B}^0\to D\bar{K}^{*0}$ are obtained with the substitution
$-\phi_3 \to \phi_3$ and $A \leftrightarrow \bar{A}$;
the corresponding parameters that depend on the $\bar{B}^0$ decay amplitude are
\begin{eqnarray}
x_- =r_S\cos(\delta_S - \phi_3), \hspace{20pt} y_- = r_S\sin(\delta_S - \phi_3).
	\label{eq:x_y}
\end{eqnarray}
If $P$, $\bar{P}$, $C$, $S$ and $k$ are known, one can obtain ($x_+$, $y_+$) from $B^0$ and ($x_-$, $y_-$) from $\bar{B}^0$ decays.
Combining both $B^0$ and $\bar{B^0}$ measurements, $r_S$, $\phi_3$ and $\delta_S$ can be extracted.

\begin{figure}[ht]
  \begin{center}
    {\includegraphics[width=0.60\textwidth]{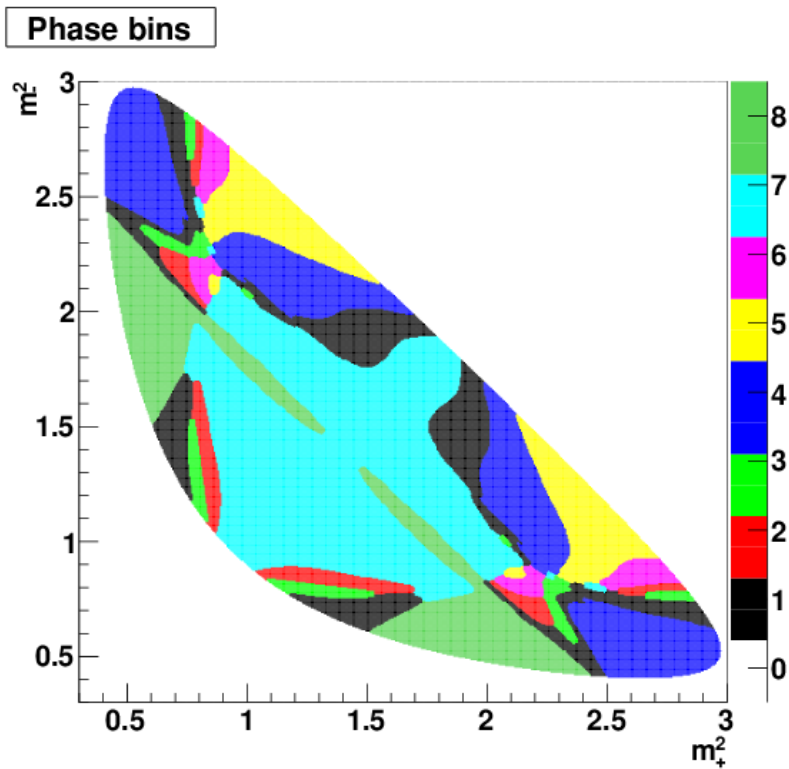}}
    \caption{
    Binning of the optimal binning in Ref.~\cite{cisi_CLEO_1, cisi_CLEO_2}.
    }
    \label{fig:bin_opt}
  \end{center}
\end{figure}

In the model-dependent analysis, one deals directly with the Dalitz distribution density and the functions $C$ and $S$ are obtained from a model based upon a fit to the $D^0\to K_S^0\pi^+\pi^-$ amplitude.
On the other hand, in the model-independent approach~\cite{MIDalitz_phi3}, where the assumption of a model for $D^0\to K_S^0\pi^+\pi^-$ decay is not necessary,
the Dalitz plot is divided into $2{\cal N}$ bins symmetric under the exchange $m_-^2 \leftrightarrow m_+^2$.
The bin index $i$ ranges from $-{\cal N}$ to ${\cal N}$ (excluding $0$);
the exchange $m_-^2 \leftrightarrow m_+^2$ corresponds to the exchange $i \leftrightarrow -i$.
The expected number of signal events in bin $i$ of the Dalitz distribution of the $D$ mesons from $B^0\to DK^{*0}$ is
\begin{eqnarray}
N^{\pm}_i = h_B \left[ K_{\pm i} + r_S^2K_{\mp i} + 2k\sqrt{K_i K_{-i}}(x_{\pm}c_i \pm y_{\pm}s_i) \right],
	\label{eq:N_i}
\end{eqnarray}
where $N^{+(-)}$ stands for the number of $B^0(\bar{B}^0)$ meson decays,
$h_B^{+(-)}$ is the normalization constant
and $K_{+i}$ is the number of events in the $i^{\mathrm{th}}$ bin of
a flavor-tagged $D^0\to K_S\pi^+\pi^-$ decays measured with a sample of inclusively reconstructed $D^{*+}\to D^0\pi^+$ decays.
Equation~(\ref{eq:N_i}) can be obtained by integrating Eq.~(\ref{eq:Dcy_inf}) over the $i^{\rm th}$ bin region.
Here, $K_i\propto \int_{\mathcal{D}_i}{|A|^2\mathrm{d}\mathcal{D}},$ and
$\mathcal{D}$ represents the Dalitz plane and $\mathcal{D}_i$ is the bin over which the integration is performed.
The values of $K_i$ are measured from a sample of flavor-tagged $D^0$ mesons obtained by reconstructing $D^{*\pm}\to D\pi^\pm$ decays.
The terms $c_i$ and $s_i$ are the amplitude-weighted averages of the functions $C$ and $S$ over the bin:
\begin{eqnarray}
c_i = \frac{\int_{{\cal D}_i} \mid A\mid\mid \bar{A}\mid C{\mathrm d}{\cal D}}
{\sqrt{\int_{{\cal D}_i} \mid A\mid^2 {\mathrm d}{\cal D}\int_{{\cal D}_i} \mid \bar{A}\mid^2 {\mathrm d}{\cal D}}}.
	\label{eq:c_i}
\end{eqnarray}
The terms $s_i$ are defined similarly with $C$ substituted by $S$.
The absence of $CP$ violation in the $D$ decay implies $c_i = c_{-i}$ and $s_i = -s_{-i}$.
The values of $c_i$ and $s_i$ can be measured using quantum-correlated $D$ pairs
produced at charm-factory experiments operating at the threshold of $D\bar{D}$ pair production.
The CLEO Collaboration has reported $c_i$ and $s_i$ values
from $CP$ tagged and flavor tagged $D\bar{D}$ events data,
and this analysis is performed with the optimal binning in Refs.~\cite{cisi_CLEO_1, cisi_CLEO_2}, 
as shown in Fig.~\ref{fig:bin_opt}.
Given that $c_i$ and $s_i$ are measured and $K_i$ and $k$ are known,
Eq.~(\ref{eq:N_i}) has only three free parameters ($x$, $y$, and $h_B$) for each of $B^0$ and $\bar{B}^0$, and can be solved.
We use the values of ($c_i$, $s_i$) for the ``optimal $D^0\to K^0_S \pi^+\pi^-$ binning" reported in Table X\hspace{-.1em}V\hspace{-.1em}I of Ref.~\cite{cisi_CLEO_2},
``the optimal binning" $K_i$ values reported in Table I\hspace{-.1em}I of Ref.~\cite{Antn_Mod_Ind_Dalitz},
and $k = 0.95 \pm 0.03$~\cite{BaBar_k}.
We have neglected charm-mixing effects in $D$ decays from both the $B^0\to DK^{*0}$ process
and in the quantum-correlated $D\bar{D}$ production~\cite{DDmix}.

\section{EVENT RECONSTRUCTION AND SELECTION}
This analysis is based on a data sample that contains $711~{\rm fb}^{-1}$ corresponding to $772~\times 10^6~B\bar{B}$ pairs,
collected  with the Belle detector at the KEKB asymmetric-energy $e^+e^-$ ($3.5$ on $8~{\rm GeV}$) collider~\cite{KEKB}
operating at the $\Upsilon(4S)$ resonance.
The Belle detector is a large-solid-angle magnetic spectrometer that consists of a silicon vertex detector,
a $50$-layer central drift chamber (CDC),
an array of aerogel threshold Cherenkov counters (ACC),
a barrel-like arrangement of time-of-flight scintillation counters (TOF),
and an electromagnetic calorimeter comprised of CsI(Tl) crystals located inside a superconducting solenoid coil that provides a $1.5~{\rm T}$ magnetic field.
An iron flux-return located outside of the coil is instrumented to detect $K_L^0$ mesons and to identify muons.
The detector is described in detail elsewhere~\cite{Belle}.

We reconstruct $B^0\to DK^{*0}$ events with $K^{*0}\to K^+\pi^-$ and $D\to K_S^0\pi^+\pi^-$.
The event selection described below is developed from studies of
continuum data taken at center-of-mass energies just below the $\Upsilon(4S)$ resonance and Monte Carlo (MC) simulated events.

The $K_S^0$ candidates are identified using the output of a neural network. Inputs to the network for a pair of oppositely-charged pions are the invariant mass, 20 kinematic parameters and particle identification (PID) information  from the ACC, TOF and the ionization energy loss in the CDC.
The $K_S^0$ selection has a simulated purity of 92.2\% and an efficiency of 75.1\%.
Charged kaon and pion candidates are identified using PID information.
The efficiency is 80-90\%
and the probability of misidentification is 6-10\%,
depending upon the momentum of hadrons and obtained using dedicated data control samples.
We reconstruct neutral $D$ mesons by combining a $K_S^0$ candidate with a pair of oppositely-charged pion candidates.
We require that the invariant mass be within $\pm15~{\rm MeV}/c^2$ ($\pm 3 \sigma$) of the nominal $D^0$ mass.
$K^{*0}$ candidates are reconstructed from $K^+\pi^-$ pairs.
We require that the invariant mass be within $\pm$50~${\rm MeV}/c^2$ of the nominal $K^{*0}$ mass.
We combine $D$ and $K^{*0}$ candidates to form $B^0$ mesons.
Candidate events are identified by the energy difference $\Delta E \equiv \sum_{i}E_i - E_{\mathrm b}$
and the beam-constrained mass $M_{\rm bc} c^2 \equiv \sqrt{E_{\mathrm b}^2 - \mid c \sum_{i}\vec{p}_i\mid^2}$,
where $E_{\mathrm b}$ is the beam energy and $\vec{p}_i$ and $E_i$ are the momenta and energies, respectively,
of the $B^0$ meson decay products in the $e^+e^-$ center-of-mass (CM) frame.
We select events with $5.21~{\rm GeV}/c^2 < M_{\rm bc} < 5.29~{\rm GeV}/c^2$ and $-0.10~{\rm GeV} < \Delta E < 0.15~{\rm GeV}$.

Among other $B$ decays, the most serious background is from
$\bar{B}^0$ decaying to the same final state as $B^0\to DK^{*0}$.
To suppress this background,
we exclude candidates for which the invariant mass of the $K^{*0} \pi^+$ system is within $\pm4~{\rm MeV}/c^2$ of the nominal $D^+$ mass.
This criterion leads to a negligible contamination from this mode
and a relative loss of 0.6\% in the signal efficiency.

The large combinatorial background of true $D^0$ and random $K^+$ and $\pi^-$ combinations
from the $e^+e^-\to c\bar{c}$ process and other $B\bar{B}$ decays
is reduced if $D^0$ candidates that are a decay product of $D^{*+} \to D^0 \pi^+$ are eliminated.
We use the mass difference $\Delta M$ between the $[K_S^0\pi^+\pi^-]_D \pi^+$ and $[K_S^0\pi^+\pi^-]_D$ systems for this purpose:
if $\Delta M > 0.15~{\rm GeV}/c^2$ for any additional $\pi^+$ candidate not used in the $B$ candidate reconstruction, the event is retained.
This requirement removes 19\% of $c\bar{c}$ background and 11\% of $B\bar{B}$ background according to MC simulation.
The relative loss in signal efficiency is 5.5\%.

In the rare case where there are multiple candidates in an event,
the candidate with $M_{\rm bc}$ closest to the nominal value is chosen.
The relative loss in signal efficiency is 0.8\%.

To discriminate signal events from the large combinatorial background dominated by the two-jet-like $e^+e^-\rightarrow q\bar{q}$ continuum process,
where $q$ indicates $u$, $d$, $s$ or $c$, a multivariate analysis is performed using the 12 variables introduced in Table~\ref{tab:paras}.
\begin{table}
\begin{center}
\begin{tabular}{| c | l |}
\hline
1 & \parbox{420pt}{\strut{} 
Fisher discriminants based on modified Fox-Wolfram moments~\cite{SFW}.
\strut} \\
\hline
2 & \parbox{420pt}{\strut{}
 The angle in the CM frame between the thrust axes of the $B$ decay
 and that of remaining particles.
\strut} \\
\hline
3 & \parbox{420pt}{\strut{}
The signed difference of the vertices between the $B$ candidate and the remaining charged tracks.
\strut} \\
\hline
4 & \parbox{420pt}{\strut{}
The distance of closest approach between the trajectories of the $K^{*}$ and $D$ candidates.
\strut} \\
\hline
5 & \parbox{420pt}{\strut{}
The expected flavor dilution factor described in Ref.~\cite{TaggingNIM}.
\strut} \\
\hline
6 & \parbox{420pt}{\strut{}
The angle $\theta$ between the $B$ meson momentum direction and the beam axis in the CM frame.
\strut} \\
\hline
7 & \parbox{420pt}{\strut{}
The angle between the $D$ and $\Upsilon(4S)$ directions in the rest frame of the $B$ candidate.
\strut} \\
\hline
8 & \parbox{420pt}{\strut{}
The projection of the sphericity vector with the largest eigenvalue onto the $e^+e^-$ beam direction.
\strut} \\
\hline
9 & \parbox{420pt}{\strut{}
The angle of the sphericity vector with the largest eigenvalue with respect to that of the remaining particles.
\strut} \\
\hline
10 & \parbox{420pt}{\strut{}
The angle of the sphericity vector with the second largest eigenvalue.
\strut} \\
\hline
11 & \parbox{420pt}{\strut{}
The angle of the sphericity vector with the smallest eigenvalue.
\strut} \\
\hline
12 & \parbox{420pt}{\strut{}
The magnitude of the thrust of the particles not used to reconstruct the signal.
\strut} \\
\hline
\end{tabular}
\caption{
Variables used for $q\bar{q}$ suppression.
}
\label{tab:paras}
\end{center}
\end{table}
To effectively combine these 12 variables, we employ the NeuroBayes neural network package \cite{NB}.
The NeuroBayes output is denoted as $C_{\rm NB}$ and lies within the range  [$-1$, 1];
events with $C_{\rm NB} \sim 1$ are signal-like and events with $C_{\rm NB} \sim -1$ are $q\bar{q}$-like.
Training of the neural network is performed using signal and $q\bar{q}$ MC samples.
The $C_{\rm NB}$ distribution of signal events peaks at $C_{\rm NB} \sim 1$ and is therefore difficult to represent with a simple analytic function.
However, the transformed variable
\begin{eqnarray}
  C'_{\rm NB} &=& \ln \frac{C_{\rm NB}-{C_{{\rm NB,} {\rm low}}}}{C_{{\rm NB,} {\rm high}}-C_{\rm NB}}\ ,
\end{eqnarray}
where $C_{{\rm NB,} {\rm low}} = -0.6$ and $C_{{\rm NB,} {\rm high}} = 0.9992$,
has a distribution that can be modeled by a Gaussian for signal as well as background.
The events with $C_{\rm NB} < C_{{\rm NB,} {\rm low}}$ are rejected;
the relative loss in signal efficiency is 7.4\%.

\section{ANALYSIS PROCEDURE}
In this section we describe the fit to determine the physics parameters.
In Sec.~\ref{sec:bg_sp} we describe the signal and background shape parametrization.
In Sec.~\ref{sec:bin_by_bin} we describe how we correct for the effect of migration and acceptance variations between bins.
In Sec.~\ref{seq:xy_fit} the fit to extract the values of ($x$, $y$) is described.

\subsection{Signal and background parametrization}
\label{sec:bg_sp}
The number of signal events is obtained by fitting the three-dimensional distribution of variables $M_{\rm bc}$, $\Delta E$, and $C'_{\rm NB}$
using the extended maximum likelihood method.
We form three-dimensional PDFs for each component as the product of one-dimensional PDFs for $\Delta E$, $M_{\rm bc}$ and $C'_{\rm NB}$,
since the correlations among the variables are found to be small.
The fit region is defined as $\Delta E\in[-0.1, 0.15]~{\rm GeV}$ and $M_{\rm bc} > 5.21~{\rm GeV}/c^{2}$.

Backgrounds are divided into the following components:
\begin{itemize}
\item	Continuum background from $q\bar{q}$ events.
\item	$B\bar{B}$ background, in which the tracks forming the $B^0\to DK^{*0}$ candidate come from decays of both $B$ mesons in the event.
	The number of possible $B$ decay combinations that contribute to this background is large;
	therefore, both the Dalitz distribution and distribution of the fit parameters are quite smooth.
	$B\bar{B}$ backgrounds are further subdivided into two components:
	events reconstructed with a true $D\to K^{0}_{S}\pi^{+}\pi^{-}$ decay, referred to as $D_{\rm true}$ $B\bar{B}$ background,
	and those reconstructed with a combinatorial $D$ candidate, referred to as $D_{\rm fake}$ $B\bar{B}$ background.
\item	Peaking $B\bar{B}$ background, in which all tracks forming the $B^0\to DK^{*0}$ candidate arise from the same $B$ meson.
	This background has two types: events with one pion misidentified as a kaon, such as $D^0 [\pi^+\pi^-]_{\rho^0}$, and
	one pion misidentified as a kaon and one pion not reconstructed, such as $D^0 [\pi^+\pi^+\pi^-]_{a_1^+}$.
	The backgrounds come from individual $B$ decays and are well separated from the signal.
\end{itemize}

The $\Delta E$ PDFs are parameterized by
a double Gaussian for the signal,
an exponential function for the $D_{\rm true}$ $B\bar{B}$ background,
an exponential function for the $D_{\rm fake}$ $B\bar{B}$ background,
a linear function for the $q\bar{q}$ background,
a double Gaussian for the $\bar{D}^0\rho^0$ background,
and a Gaussian for the $\bar{D}^0a_1^+$ background.
The $M_{\rm bc}$ PDFs are
a Gaussian for signal,
a Crystal Ball function~\cite{Crystall-Ball} for $D_{\rm true}$ $B\bar{B}$ background,
an ARGUS function~\cite{ARGUS} for $D_{\rm fake}$ $B\bar{B}$ background,
an ARGUS function for $q\bar{q}$ background,
a sum of a Gaussian and ARGUS functions for $\bar{D}^0\rho^0$ background
and a Gaussian for $\bar{D}^0a_1^+$ background.
For each component, the $C'_{\rm NB}$ PDF is the sum of a Gaussian and bifurcated Gaussian.
The shape parameters of the  PDFs are fixed from MC samples.

\begin{figure}[]
  \begin{center}
  \includegraphics[width=1.0\textwidth]{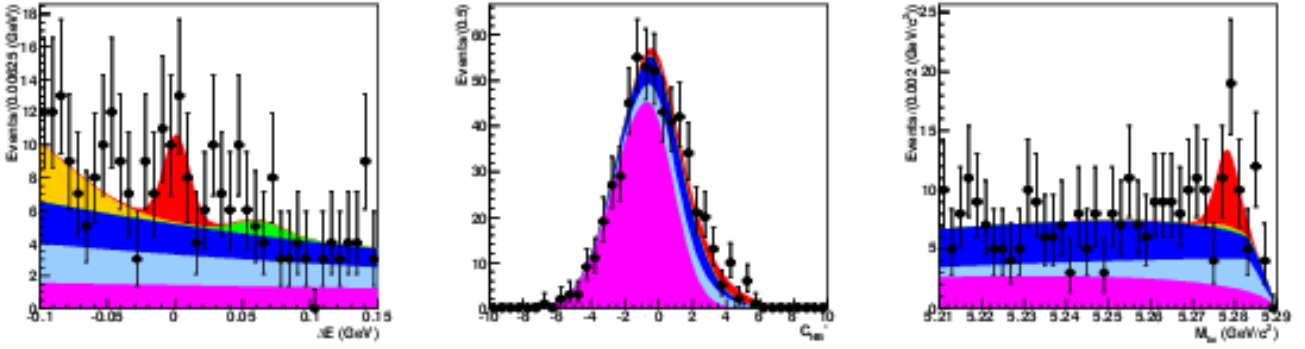}
      \caption{Projection of the fit to real data using the full Dalitz plot.
	Left: $\Delta E$ distribution with $M_{\rm bc} > 5.27~{\rm GeV}/c^{2}$ and $C'_{\rm NB} > 2$ requirements.
    	Middle: $C'_{\rm NB}$ distribution with $\mid\Delta E\mid < 0.03~{\rm GeV}$ and $M_{\rm bc} > 5.27~{\rm GeV}/c^{2}$ requirements.
    	Right: $M_{\rm bc}$ distribution with $\mid\Delta E\mid < 0.03~{\rm GeV}$ and $C'_{\rm NB} > 2$ requirements.
	Curves show the fitted signal and background contributions,
	(red is signal,
	yellow is $D^0a_1^+$,
	green is $D^0\rho^0$,
	blue is $D_{\rm fake}$ $B\bar{B}$,
	light blue is $D_{\rm true}$ $B\bar{B}$ and
	magenta is $q\bar{q}$)
	and points with error bars are the data.
	}
    \label{fig:fit_data}
  \end{center}
\end{figure}

The numbers of events in each bin are free parameters in the fit.
This procedure has been justified for background that is either well separated from the signal
(such as peaking $B\bar{B}$ background)
or is constrained by a much larger number of events than the signal
(such as $q\bar{q}$ background).
The results of the fit to the full Dalitz plot are shown in Fig.~\ref{fig:fit_data}.
We obtain a total of $44.2 ^{+13.3}_{-12.1}$ signal events.
The statistical significance is $2.8 \sigma$ relative to the no-signal hypothesis.
Simultaneously, we obtain 
$695.8^{+177.6}_{-175.6}$ for $D_{\rm true}$ $B\bar{B}$,
$1963.2^{+228.1}_{-227.5}$ for $D_{\rm fake}$ $B\bar{B}$,
$11075.7^{+156.6}_{-155.5}$ for $q\bar{q}$,
$16.6^{+16.7}_{-13.6}$ for $\bar{D}^0\rho^0$ and
$59.3^{+22.3}_{-20.8}$ for $\bar{D}^0a_1^+$ backgrounds events.

\subsection{Corrections to the bin-by-bin yields}
\label{sec:bin_by_bin}
There are further effects that must be accounted for before the values of ($x$, $y$) can be determined from a binned fit.
Equation~	\ref{eq:N_i} only holds if there is no migration between bins and the Dalitz acceptance is uniform.
Here, we consider the crossfeed for bin-by-bin yields as the migration and the acceptance as the event reconstruction efficiency.

First, we discuss migration which is due to momentum resolution and flavor misidentification.
Momentum resolution leads to migration of events among the bins.
In the binned approach, this effect can be corrected in a non-parametric way.
The migration can be described by a linear transformation of the number of events in each bin
\begin{eqnarray}
N_{{\rm obs,}i} = \sum \alpha_{ik}N_{k}^\prime,
\end{eqnarray}
where $N^\prime_i$ is the 
the number of events that bin $i$ would contain without the migration with acceptance
and $N_{{\rm obs,}i}$ is the reconstructed number of events in bin $i$.
The migration matrix $\alpha_{ik}$ is nearly the unit matrix;
it is obtained from a signal MC simulation generated with the amplitude model reported in Ref.~\cite{Dalitz_Belle_result}.
Most of the off-diagonal elements are null; only a few have values $|\alpha_{ik}| \le 0.04$. 
In the case of a $D\to K_S^0\pi^+\pi^-$ decay from a $B$, the migration depends on the parameters $x$ and $y$.
However, this is a minor correction to an already small effect and so is neglected.

The second migration effect to be considered is due to misidentification of the $B$ flavor.
Double misidentification in $K^{*0}$ reconstruction from $K^+\pi^-$
where $K^-$ is misidentified as $\pi^-$ and $\pi^+$ is misidentified as $K^+$ at the same time
leads to migration of events between $N_i^+ \leftrightarrow N_{-i}^-$
due to assignment of the wrong flavor to the $B$ candidate.
If the fraction of doubly-misidentified events is $\beta$, the number of events in each bin can be  written as
\begin{eqnarray}
N_{i}^{\prime \pm} = N_{{\rm obs,}i}^{\pm} + \beta N_{{\rm obs,}-i}^{\mp}.
\end{eqnarray}
The value of $\beta$ is obtained from  MC simulation and is found to be $(0.12 \pm 0.01)\%$.
Therefore, the effect of flavor misidentification is neglected.

The final state radiation also causes migration between bins.
The measured values of $c_i$ and $s_i$ by CLEO are not corrected for the radiation and the effect upon our analysis is found to be negligible \cite{Antn_Mod_Ind_Dalitz}.

Second, we consider the effect of the variation of the efficiency profile over the Dalitz plane.
We note that Eq.~(\ref{eq:Dcy_inf}) does not change under the transformation $P\to \epsilon P$ when the efficiency profile
$\epsilon(m_+^2, m_-^2)$ is symmetric:
$\epsilon(m_+^2, m_-^2) = \epsilon(m_-^2, m_+^2)$.
The effect of non-uniform efficiency over the Dalitz plane cancels
when using a flavor-tagged $D$ sample with kinematic properties that are similar to the sample from the signal $B$ decay.
This approach allows for the removal of the systematic uncertainty associated
with the possible inaccuracy of the detector acceptance description in the MC simulation.
With the efficiency taken into account (that is, in general non-uniform across the bin region),
the number of events reconstructed is
\begin{eqnarray}
N^{\prime} = \int p({\cal D})\epsilon({\cal D}){\mathrm d}{\cal D}.
	\label{eq:N_i_eff}
\end{eqnarray}
Here, $p$ is the probability density on the Dalitz plane and ${\cal D}$ is the position on Dalitz plane.
Clearly, the efficiency does not factorize.
One can use an efficiency averaged over the bin, then correct for it in the analysis:
\begin{eqnarray}
\bar{\epsilon}_i = \frac{N_i^\prime}{N_i} = \frac{\int p({\cal D})\epsilon({\cal D}){\mathrm d}{\cal D}}{\int p({\cal D}){\mathrm d}{\cal D}}.
	\label{eq:eff}
\end{eqnarray}
Here, $N_i$ are the number of events corrected for variations in acceptance and migration,
which should be used for ($x_\pm$, $y_\pm$) extraction.
The averaged efficiency $\bar{\epsilon}_i$ can be determined from MC.
The assumption that the efficiency profile depends only on the $D$ momentum is tested using MC simulation
and the residual difference is treated as a systematic uncertainty.
The correction for $c_i$ and $s_i$ due to efficiency variation within a bin cannot be calculated in a completely model-independent way,
since the correction terms include the amplitude  variation inside the bin.
Calculations using the Belle $D\to K_S^0\pi^+\pi^-$ model~\cite{Dalitz_Belle_result} show
that this correction is negligible even for very large non-uniformity of the efficiency profile.

\subsection{Fit to determine ($x$, $y$)}
\label{seq:xy_fit}
If $N_i$ in each bin is measured, $x_\pm$ and $y_\pm$
can be obtained according to Eq.~(\ref{eq:N_i}) by minimizing
\begin{eqnarray}
-2\log{\cal L}(x, y) = -2\sum_i \log p(\left< N_i \right>(x, y), N_i, \sigma_{N_i}),
\label{eq:NLL}
\end{eqnarray}
where $\left< N_i \right>$ are the expected number of signal events in the bin $i$ obtained from Eq.~(\ref{eq:N_i}).
Here, $N_i$ and $\sigma_{N_i}$ are the observed number of events in data and the uncertainty on $N_i$, respectively.

The procedure described above does not make any assumptions about the Dalitz distribution of the background events,
since the fits in each bin are independent.
Thus, there is no uncertainty related to the Dalitz model.
However, in our case, where there are a small number of events and many background components, such independent fits are not feasible.
Therefore, we obtain ($x_{\pm}$, $y_{\pm}$) from a combined fit with a common likelihood for all bins.
The relative numbers of background events in each bin are constrained to the numbers found in the MC.
The amount of the $D_{\rm true}$ $B\bar{B}$ background in bins from the ratio of $D^0$ $(K_i)$ and $\bar{D}^{0}$ $(K_{-i})$ from MC
and the amount of the $D_{\rm fake}$, $B\bar{B}$, $q\bar{q}$ and the background from individual $B$ decays from the MC.
The yields integrated over the Dalitz plot of the background components are additional free parameters.
Thus, the variables ($x_\pm$, $y_\pm$) become free parameters of the combined likelihood fit
and the assumption that the signal yield obeys a Gaussian distribution is not needed.
While the normalization parameter $h_B$ is also a free parameter of the fit,
we do not mention it in the following as it is not a quantity of interest.

\section{COMBINED FITS TO DATA}
\label{sec:data_fit_in_bins}
\begin{figure}[]
  \begin{center}
 \includegraphics[width=1.0\textwidth]{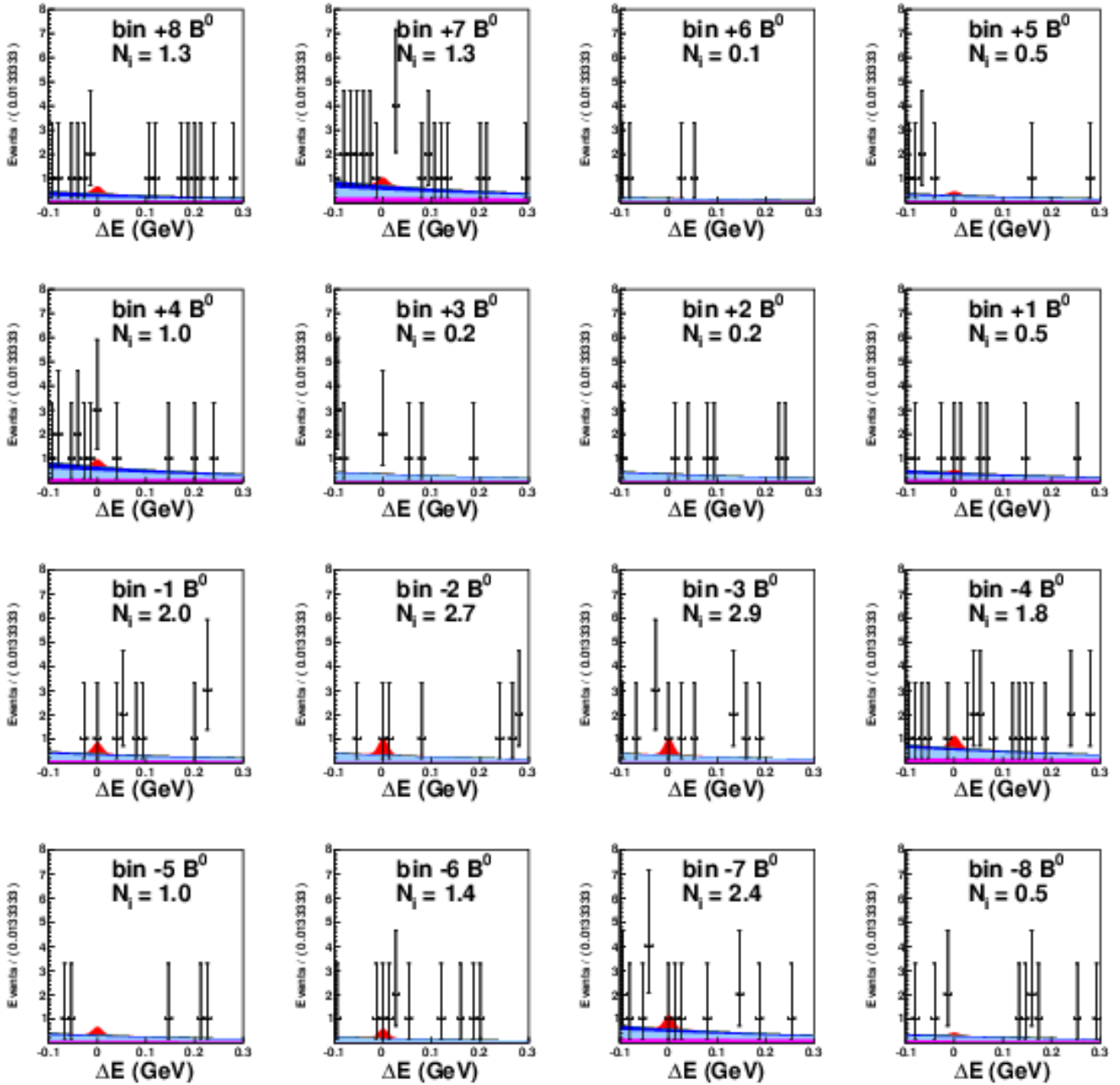}
    \caption{Projections of the combined fit of the $B^0\to DK^{*0}$ sample on $\Delta E$ for each Dalitz bin,
    	with the $M_{\rm bc} > 5.27~{\rm GeV}/c^2$ and $C'_{\rm NB} > 2$ requirements.
	The fill styles for the signal and background components are the same as in Fig.~\ref{fig:fit_data}.
	}
    \label{fig:B0_binned}
  \end{center}
\end{figure}

\begin{figure}[]
  \begin{center}
 \includegraphics[width=1.0\textwidth]{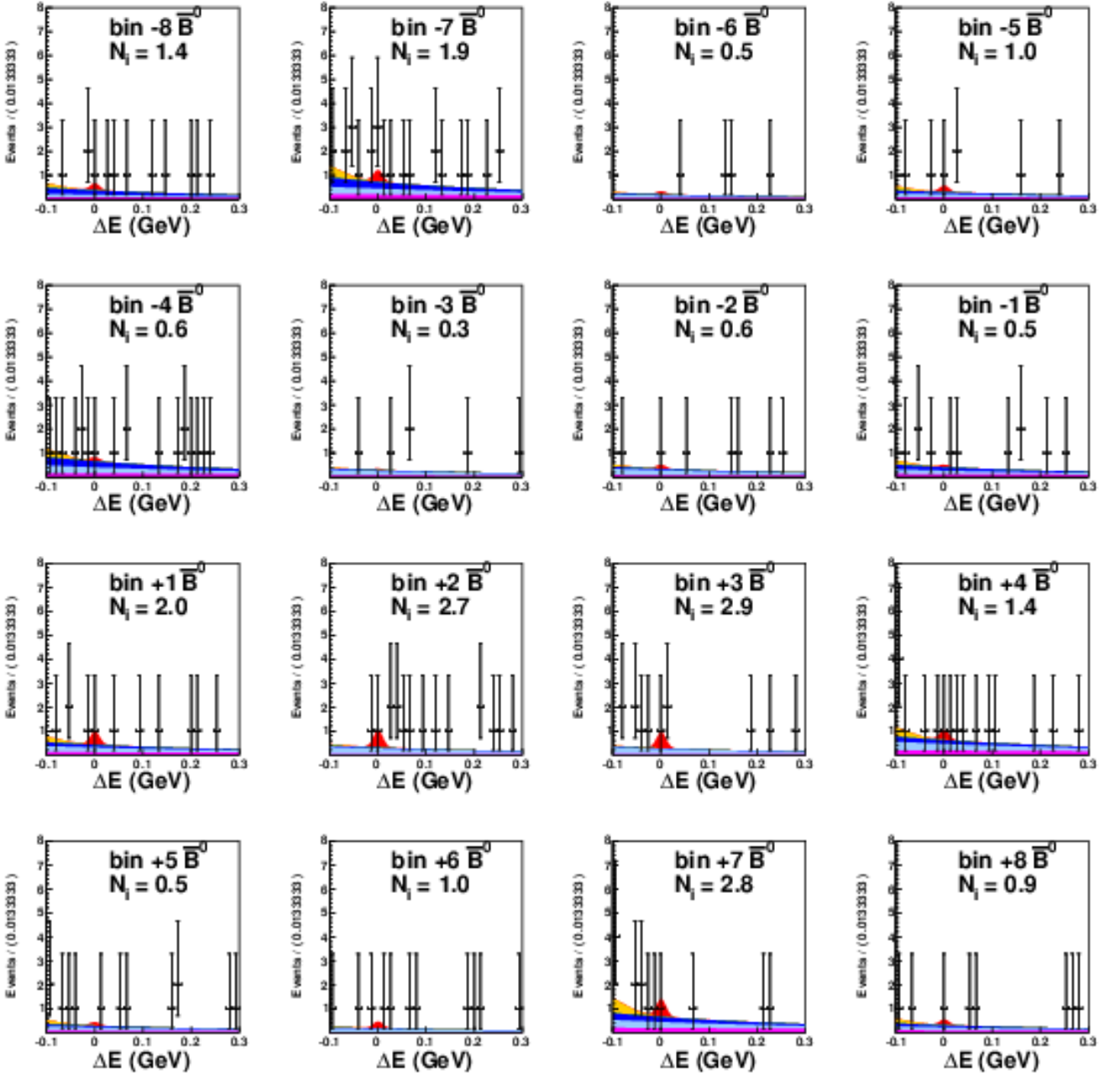}
    \caption{Projections of the combined fit of the $\bar{B}^0\to D\bar{K}^{*0}$ sample on $\Delta E$ for each Dalitz bin,
    	with the $M_{\rm bc} > 5.27~{\rm GeV}/c^2$ and $C'_{\rm NB} > 2$ requirements.
	The fill styles for the signal and background components are the same as in Fig.~\ref{fig:fit_data}.
	}
    \label{fig:B0b_binned}
  \end{center}
\end{figure}

The results of the combined fit in each bin of the $B^{0}$ and $\bar{B}^{0}$ are shown in Figs.~\ref{fig:B0_binned} and \ref{fig:B0b_binned}, respectively.
The plots show the projections of the data and the fitting model on the $\Delta E$ variable,
with the additional requirements $M_{\rm bc} > 5.27~{\rm GeV}/c^2$ and $C'_{\rm NB} > 2$.
The values of the ($x$, $y$) parameters and their statistical correlations, obtained from the combined fit for the signal sample, are given in Table~\ref{tab:xy_data}.
In this study, these ($x$, $y$) values from the likelihood distribution of the combined fit are corrected
using the frequentist approach with Feldman-Cousins ordering~\cite{FC}, which is described in Sec.~\ref{sec:result_xy_rS}.
\begin{table}
	\begin{center}
	\begin{tabular}{|c|c|}
	\hline
Parameter		& 					\\
\hline
$x_-$		& $+0.29 \pm 0.32$		\\
$y_-$		& $-0.33 \pm 0.41$		\\
corr.($x_-$, $y_-$)	& $+7.0\%$		\\
\hline
$x_+$		& $+0.07 \pm 0.42$		\\
$y_+$		& $+0.05 \pm 0.45$		\\
corr.($x_+$, $y_+$)	& $-7.5\%$		\\
\hline
	\end{tabular}
	\caption{($x$, $y$) parameters and their statistical correlations from the combined fit of the $B^0\to DK^{*0}$ sample.
		The error is statistical.
		The values and errors are obtained from the likelihood distribution.
		}
	\label{tab:xy_data}
	\end{center}
\end{table}

\section{SYSTEMATIC UNCERTAINTIES}
The systematic uncertainties of ($x$, $y$) are obtained by taking deviation from the default procedure under various assumptions.
\begin{table}[htbp]
  \begin{center}
    \begin{tabular}{|l|cccc|}
      \hline
	Source of uncertainty 	& $\Delta x_-$ 			& $\Delta y_-$ 			& $\Delta x_+$ 			& $\Delta y_+$ 			\\
      \hline
	Dalitz efficiency 		& $\pm0.00$ 			& $^{+0.01}_{-0.00}$ 	& $\pm0.01$ 			& $^{+0.00}_{-0.01}$ 	\\
	Migration between bins 	& $\pm0.00$ 			& $^{+0.01}_{-0.00}$ 	& $^{+0.01}_{-0.00}$ 	& $\pm0.00$ 			\\
	PDF parameterization	& $^{+0.01}_{-0.07}$ 	& $^{+0.07}_{-0.01}$ 	& $^{+0.01}_{-0.10}$ 	& $^{+0.04}_{-0.06}$ 	\\
	Flavor-tag statistics	 	& $\pm0.00$ 			& $\pm0.00$ 			& $\pm0.00$ 			& $^{+0.00}_{-0.01}$ 	\\
	$c_i$, $s_i$ precision 	& $\pm0.03$ 			& $^{+0.09}_{-0.08}$ 	& $\pm0.05$ 			& $^{+0.08}_{-0.10}$ 	\\
	$k$ precision 			& $\pm0.00$ 			& $\pm0.01$ 			& $\pm0.00$ 			& $\pm0.00$ 			\\
      \hline
	Total without $c_i$, $s_i$ precision 	& $^{+0.01}_{-0.07}$ 	& $^{+0.07}_{-0.02}$ 	& $^{+0.02}_{-0.10}$ 	& $^{+0.04}_{-0.06}$ 	\\
      \hline
	Total 						& $^{+0.03}_{-0.08}$ 	& $^{+0.12}_{-0.08}$ 	& $^{+0.05}_{-0.11}$ 	& $^{+0.09}_{-0.12}$ 	\\
      \hline
       \end{tabular}
       \caption{Systematic uncertainties in the ($x$, $y$) measurement for the $B^0\to DK^{*0}$ mode.
Values are rounded to two significant digits and those less than 0.005 are quoted as 0.00.
}
    \label{tab:syst_err}
  \end{center}
\end{table}
The systematic uncertainties are summarized in Table~\ref{tab:syst_err};
most are negligible compared to the statistical uncertainty.
There is an uncertainty due to the Dalitz efficiency variation
because of the difference in average efficiency over each bin
for the flavor-tagged $D$ and $B^0\to DK^{*0}$ samples.
A maximum difference of 1.5\% is obtained in a MC study.
The uncertainty is taken as the maximum of two quantities:
\begin{itemize}
\item the root mean square of $x$ and $y$ from smearing the numbers of events in the flavor-tagged sample $K_i$ by 1.5\%, or
\item the bias in $x$ and $y$ between the fits with and without efficiency correction for $K_i$ obtained from signal MC.
\end{itemize}
The uncertainty due to migration of events between bins is estimated by taking the bias between the fits with and without the migration correction.
The uncertainties due to the fixed parameterization of the signal and background PDFs are estimated by varying them by $\pm 1 \sigma$.
The uncertainty due to the $C'_{\rm NB}$ PDF distributions for ${B\bar{B}}$ is estimated by replacing them with the signal $C'_{\rm NB}$ PDF.
The uncertainty due to the $D_{\rm true}$ and $D_{\rm fake}$ $B\bar{B}$ fractions is estimated by varying them between 0 and 1.
The uncertainty arising from the finite sample of flavor-tagged $D\to K_S^0\pi\pi$ decays
is evaluated by varying the values of $K_i$ within their statistical uncertainties.
The uncertainty due to the limited precision of $c_i$ and $s_i$ parameters
is obtained by smearing the $c_i$ and $s_i$ values within their total errors and repeating the fits for the same experimental data.
The uncertainty due to $k$ in Eq.~(\ref{eq:rS_delS_k}) is evaluated by varying the value of $k(=0.95\pm0.03)$ within its error~\cite{BaBar_k}.
Total systematic uncertainties in the ($x$, $y$) are obtained by summing all uncertainties in quadrature and listed in Table \ref{tab:syst_err}.

\section{RESULT}
\label{sec:result_xy_rS}
We use the frequentist approach with Feldman-Cousins ordering~\cite{FC}
to obtain the physical parameters $\mu = (\phi_3, r_S, \delta_S)$ (or true parameters $\mu = z_{\rm true} = (x_-, y_-, x_+, y_+$))
from the measured parameters $z = z_{\rm meas} = (x_-, y_-, x_+, y_+)$ taken from the likelihood distribution.
In essence, the confidence level $\alpha$ for a set of physical parameters $\mu$ is calculated as
\begin{eqnarray}
\alpha(\mu) = \frac{\int_{{\cal D}(\mu)}p(z\mid \mu){\mathrm d}z}{\int_{\infty}p(z\mid \mu){\mathrm d}z},
	\label{eq:CL}
\end{eqnarray}
where $p(z\mid \mu)$ is the probability density to obtain the measurement $z$ given by the set of true parameters $\mu$.
The integration domain ${\cal D}(\mu)$ is given by the likelihood ratio (Feldman-Cousins) ordering:
\begin{eqnarray}
\frac{p(z\mid\mu)}{p(z\mid\mu_{\rm best}(z))} > \frac{p(z_0\mid\mu)}{p(z_0\mid\mu_{\rm best}(z_0))},
	\label{eq:int_reg}
\end{eqnarray}
where $\mu_{\rm best}(z)$ is $\mu$ that maximizes $p(z\mid\mu)$ for the given $z$, and $z_0$ is the result of the data fit.
This PDF is taken from MC pseudo-experiments.

Systematic uncertainties in $\mu$ are obtained by varying the measured parameters $z$ within their systematic uncertainties assuming a nominal distribution.
In this calculation, we ignore the correlations of
uncertainties between the $B^0$ and $\bar{B}^0$ as the two samples are independent.

\begin{figure}[]
  \begin{center}
	\includegraphics[width=0.56\textwidth]{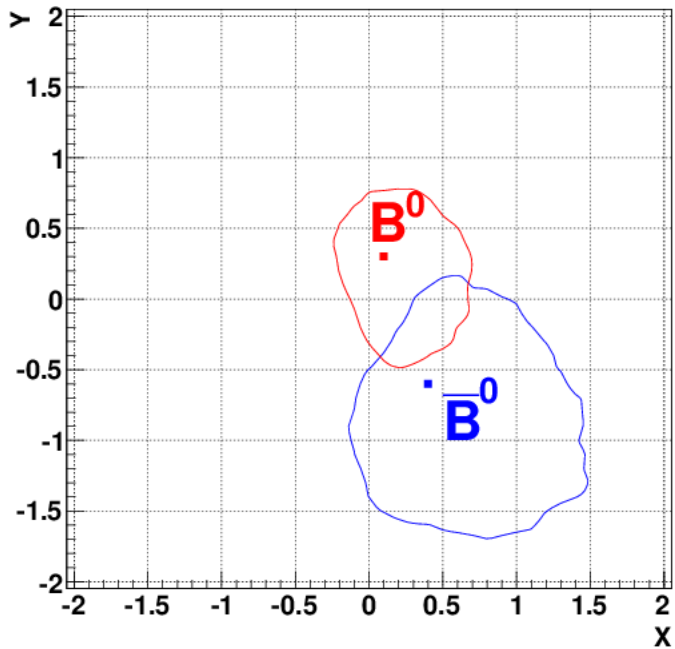}
    		\caption{C.L. contours for ($x_-$, $y_-$) (blue) and ($x_+$, $y_+$) (red).
    			The dots show the most probable ($x$, $y$) values;
			the lines show the 68\% contours.
			The fluctuations arise from the statistics of the pseudo-experiments and C.L. step used.
	    	}	
    \label{fig:xy_CL}
  \end{center}
\end{figure}

\begin{figure}[]
  \begin{center}
    \includegraphics[width=0.7\textwidth]{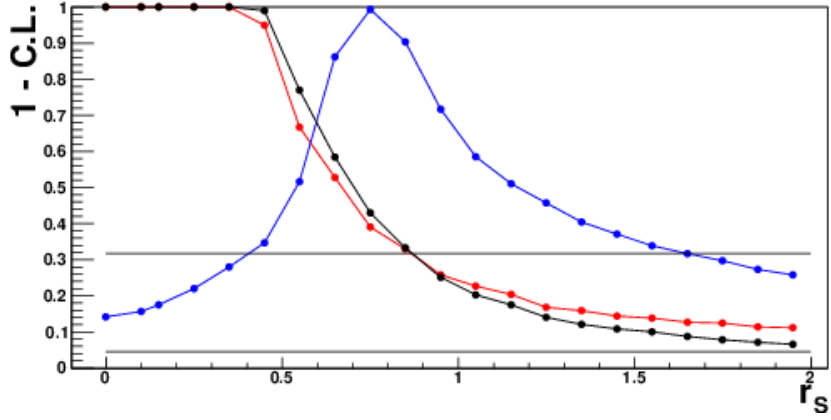}
    \caption{Likelihood profile for $r_S$.
	The blue points are for  $\bar{B}^0$ ($x_-$, $y_-$),
	red are for $B^0$ ($x_+$, $y_+$)
	and black are $\bar{B}^{0}$ and $B^{0}$ combined.
	The two horizontal lines show 68\% and 95\% C.L.
    	}
    \label{fig:rS_CL}
  \end{center}
\end{figure}

As a result of this procedure, we obtain the confidence levels (C.L.) for ($x$, $y$) and the physical parameter $r_S$.
The C.L. contours on ($x$, $y$) are shown in Fig.~\ref{fig:xy_CL}.
The $1-\mathrm{C.L.}$ as a function of $r_S$ is shown in Fig.~\ref{fig:rS_CL}.
The final results are:
\begin{eqnarray}
x_- &=& +0.4 ^{+1.0 +0.0}_{-0.6 -0.1} \pm0.0,	\\
y_- &=& -0.6 ^{+0.8 +0.1}_{-1.0 -0.0} \pm0.1,	\\
x_+ &=& +0.1 ^{+0.7 +0.0}_{-0.4 -0.1} \pm0.1,	\\
y_+ &=& +0.3 ^{+0.5 +0.0}_{-0.8 -0.1} \pm0.1,	\\
r_S &<& 0.87 \hspace{30pt} {\rm at}~68\% ~{\rm C.L.},
\end{eqnarray}
where the first error is statistical, the second is systematic without uncertainties in ($c_i$, $s_i$), and the third is from the ($c_i$, $s_i$) precision from CLEO.

\section{CONCLUSION}
We report the first measurement of the amplitude ratio $r_S$
using a model-independent Dalitz analysis of $D\to K_S\pi^+\pi^-$ decays in the process $B^0\to DK^{*0}$
with the full data sample of $711~{\rm fb}^{-1}$ corresponding to $772\times 10^6~B\bar{B}$ pairs
collected by the Belle detector at the $\Upsilon(4S)$  resonance.
Model independence is achieved by binning the Dalitz plot of the $D\to K^0_S\pi^+\pi^-$ decay and using the strong-phase coefficients
with binning as in the CLEO experiment~\cite{cisi_CLEO_2}.
We obtain the value $r_S < 0.87$ at 68\% C.L.
This measurement results in lower statistical precision than the model-dependent measurement
from BaBar with the $B^0\to DK^{0}$ mode~\cite{Dalitz_BaBar_result} despite the larger data sample due to the smaller $B^0\to DK^{*0}$ signal observed.
The result is consistent with the most precise $r_S$ measurement reported by the LHCb Collaboration~\cite{r_S_LHCb} of $r_S = 0.240^{+0.055}_{-0.048}$
that uses $B^0\to [K^+K^-, K^\pm\pi^\mp, \pi^+\pi^-]_DK^{*0}$ decays.
We have confirmed the feasibility of the model-independent Dalitz analysis method with neutral $B\to DK^*$.
The value of $r_S$ indicates the sensitivity of the neutral $B\to DK^{*}$ decay to $\phi_3$ because the statistical uncertainty is proportional to $1/r_S$. 
In future high statistics experiments such as Belle II and the LHCb upgrade,
this method will give a precise and model independent determination of $\phi_3$.
A more advanced double-Dalitz-plot analysis of $B^0\to DK^+\pi^-$, $D\to K_S\pi^+\pi^-$~\cite{DoubleDalitz} has been proposed;
this result can be considered as that from one bin of such an analysis.

\end{document}